\documentstyle[12pt]{article}

\begin{document}
{\sf \begin{center} \noindent
{\Large \bf Topology and existence of 3D anisotropic filamentary kinematic dynamos}\\[3mm]

by \\[0.3cm]

{\sl L.C. Garcia de Andrade}\\

\vspace{0.5cm} Departamento de F\'{\i}sica
Te\'orica -- IF -- Universidade do Estado do Rio de Janeiro-UERJ\\[-3mm]
Rua S\~ao Francisco Xavier, 524\\[-3mm]
Cep 20550-003, Maracan\~a, Rio de Janeiro, RJ, Brasil\\[-3mm]
Electronic mail address: garcia@dft.if.uerj.br\\[-3mm]
\vspace{2cm} {\bf Abstract}
\end{center}
\paragraph*{}
Curvature and helicity topological bounds for the magnetic energy of
the streamlines magnetic structures of a kinematic dynamo flow are
computed. The existence of the filament dynamos are determined by
solving the magnetohydrodynamic equations for 3D flows and the
solution is used to determine these bounds. It is shown that in the
limit of zero resistivity filamentary dynamos always exists in the
isotropic case, however when one takes into account that the Frenet
frame does not depend only of the filament length parameter s,
(anisotropic case) the existence of the filamentary dynamo structure
depends on the curvature in the case of screwed dynamos. Frenet
curvature is associated with forld and torsion to twist which allows
us to have a sretch, twist, and fold method to build fast filament
dynamos. Arnold theorem for the helicity bounds of energy of a
divergence-free vector field is satisfied for these streamlines and
the constant which depends on the size of the compact domain $M C
R^{2}$, where the vector field is defined is determined in terms of
the dimensions of the constant cross-section filament. It is shown
that when the Arnold theorem is violated by the filament
amplification of the magnetic field structure appears, the magnetic
field decays in space.\vspace{0.5cm} \noindent {\bf PACS numbers:}
\hfill\parbox[t]{13.5cm}{02.40.Hw:Riemannian geometries}

\newpage
\section{Introduction}
 The topology and geometry of hydrodynamical and MHD dynamos have been mainly developed by Arnold and Khesin
 \cite{1} and by Childress and Gilbert \cite{2} using non chaotic flows and the twist, stretch and fold technique developed
 by Moffatt \cite{3} to investigate fast dynamos. Dynamos in chaotic flows have been developed lately by Thiffault and
 Boozer \cite{4}. On the other hand twisted filamentary magnetic structures have been important in plasma and solar
 physics \cite{5} in the investigation of electric carrying-current loops. Though we know \cite{2} that planar dynamos do
 not exist they amplify the magnetic fields. It is also important to recognize when we have a dynamo or anti-dynamo from the magnetic
 profile. Since we only have planar dynamos in very specific
 circunstances, when we have folds of filaments for example, we here
 consider three dimensional 3D dynamos where the twist can be
 associated with torsion of the filament and curvature to fold
 processes. In this paper we consider the topology and geometrical bound obtained from
 the solution of magnetic MHD equations with a dissipative term in the absence of electric potential. The scalar
 MHD equations to be solved are obtained by expressing the MHD equations in the Frenet
 frame anisotropic basis where the basis depend upon not only the position of the frame along the filament but also of
 the normal and binormal direction of the frame as well as the time. The filament are considered to be planar with constant
 curvature. The existence of anti-dynamos where the magnetic field is amplifies, as in the case of galactic magnetic fields,
 depends upon the relation between the right (on the left) handness of curvature of the filament
 and the normal direction to the filament which is assumed to move along the binormal direction such as some vortex
 filaments. Actually this is not the first time Frenet curvature is used to investigate dynamos, in 2001 Schekochihin et al \cite{6}used the statistical Frenet
 curvature in the small-scale magnetic fields in kinematic dynamos. This paper is organized as follows: In section 2
 we compute the filamentary solution of the magnetic filament.
 In section 3 we compute the energy bounds from helicity and curvature and examine the relation with the existence of the
 filamentary dynamos. The conclusions are presented in section 4.
 \section{MHD scalar equations for kinematical dynamos}
 Let us now start by considering the MHD field equations
\begin{equation}
{\nabla}.\vec{B}=0 \label{1}
\end{equation}
\begin{equation}
\frac{{\partial}}{{\partial}t}\vec{B}-{\nabla}{\times}[\vec{u}{\times}\vec{B}]-{\epsilon}{\nabla}^{2}\vec{B}=0
 \label{2}
\end{equation}
where $\vec{u}$ is a solenoidal field while ${\epsilon}$ is the
diffusion coefficient. Equation (\ref{2}) represents the induction
equation. The magnetic field $\vec{B}$ is chosen to lie along the
filament and is defined by the expression $\vec{B}=B({s,n})\vec{t}$
and $\vec{u}=u\vec{b}$, chosen by the similarity with vortex
filaments, is the speed of the flow. The remaining coordinate $n$ is
orthogonal to the filament all along its extension, and the arc
length s measures distances along the the filament itself. The
vectors $\vec{t}$ and $\vec{n}$ along with binormal vector $\vec{b}$
together form the Frenet frame which obeys the Frenet-Serret
equations
\begin{equation}
\vec{t}'=\kappa\vec{n} \label{3}
\end{equation}
\begin{equation}
\vec{n}'=-\kappa\vec{t}+ {\tau}\vec{b} \label{4}
\end{equation}
\begin{equation}
\vec{b}'=-{\tau}\vec{n} \label{5}
\end{equation}
the dash represents the ordinary derivation with respect to
coordinate s, and $\kappa(s,t)$ is the curvature of the curve where
$\kappa=R^{-1}$. Here ${\tau}$ represents the Frenet torsion. We
follow the assumption that the Frenet frame \cite{7} may depend on
other degrees of freedom such as that the gradient operator becomes
\begin{equation}
{\nabla}=\vec{t}\frac{\partial}{{\partial}s}+\vec{n}\frac{\partial}{{\partial}n}+\vec{b}\frac{\partial}{{\partial}b}
\label{6}
\end{equation}
 The other equations for the other legs of the Frenet frame are
\begin{equation}
\frac{\partial}{{\partial}n}\vec{t}={\theta}_{ns}\vec{n}+[{\Omega}_{b}+{\tau}]\vec{b}
\label{7}
\end{equation}
\begin{equation}
\frac{\partial}{{\partial}n}\vec{n}=-{\theta}_{ns}\vec{t}-
(div\vec{b})\vec{b} \label{8}
\end{equation}
\begin{equation}
\frac{\partial}{{\partial}n}\vec{b}=
-[{\Omega}_{b}+{\tau}]\vec{t}-(div{\vec{b}})\vec{n}\label{9}
\end{equation}
\begin{equation}
\frac{\partial}{{\partial}b}\vec{t}={\theta}_{bs}\vec{b}-[{\Omega}_{n}+{\tau}]\vec{n}
\label{10}
\end{equation}
\begin{equation}
\frac{\partial}{{\partial}b}\vec{n}=[{\Omega}_{n}+{\tau}]\vec{t}-
\kappa+(div\vec{n})\vec{b} \label{11}
\end{equation}
\begin{equation}
\frac{\partial}{{\partial}b}\vec{b}=
-{\theta}_{bs}\vec{t}-[\kappa+(div{\vec{n}})]\vec{n}\label{12}
\end{equation}
Another set of equations which we shall need here is the time
derivative of the Frenet frame given by
\begin{equation}
\dot{\vec{t}}=[{\kappa}'\vec{b}-{\kappa}{\tau}\vec{n}] \label{13}
\end{equation}
\begin{equation}
\dot{\vec{n}}={\kappa}\tau\vec{t} \label{14}
\end{equation}
\begin{equation}
\dot{\vec{b}}=-{\kappa}' \vec{t} \label{15}
\end{equation}
A long and straithforward computation ,specially due to the
computation of ${\nabla}^{2}A$, where the vector potential
$\vec{A}=A(t,s,n)$ in principle. Substituting these equations for
the dynamics of the Frenet frame leads to the scalar MHD expressions
\begin{equation}
{\partial}_{t}A=-{\partial}_{s}{\phi}+[{{\partial}^{2}}_{n}A-A({{\theta}_{ns}}^{2}-{{\kappa}_{0}}^{2})]
\label{16}
\end{equation}
\begin{equation}
-{\kappa}{\tau}A=-uB+{\epsilon}[2{\partial}_{n}A+({\Omega}_{s}+{\tau}){\theta}_{ns}A]
\label{17}
\end{equation}
\begin{equation}
-{\theta}_{bs}A={\epsilon}[2{\partial}_{n}A{\Omega}_{s}+{\Omega}^{2}A]
\label{18}
\end{equation}
where ${\kappa}_{0}$ is the Frenet curvature of the streamlines.
These equations have already been simplified by using the relations
\begin{equation}
{\nabla}{\times}\vec{A}=\vec{B}\label{19}
\end{equation}
which yields the following differential scalar equations
\begin{equation}
{B}=-A[{\Omega}_{b}+\tau] \label{20}
\end{equation}
\begin{equation}
{\partial}_{n}A+{\kappa}A=0 \label{21}
\end{equation}
\begin{equation}
A({\Omega}_{n}+{\tau})=0 \label{22}
\end{equation}
Where the ${\Omega}'s$ represent the abnormalities of the
streamlines of the flow. Here
\begin{equation}
{\theta}_{ns}=\vec{n}.\frac{\partial}{{\partial}n}\vec{t} \label{23}
\end{equation}
When the ${\Omega}_{s}$ vanishes we note the geodesic streamlines
are obtained. As we shall see below here we are not consider
geodesic flows dynamos. By considering planar flows where torsion
vanishes and the gauge condition
\begin{equation}
{\nabla}.\vec{A}+\frac{\partial}{{\partial}t}{\phi}=0 \label{24}
\end{equation}
This equation can be expressed as
\begin{equation}
{\partial}_{s}{A}+[{\theta}_{ns}+{\theta}_{bs}]A=0 \label{25}
\end{equation}
Now by considering that A does not depend on the coordinate s this
expression reduces to
\begin{equation}
[{\theta}_{ns}+{\theta}_{bs}]A=0 \label{26}
\end{equation}
which reduces to ${\theta}_{ns}=-{\theta}_{bs}$. By making use of
this expression and the assumption that ${\phi}=0$ one simplifies
the MHD scalar equations to
\begin{equation}
{\partial}_{t}A=[{{\partial}^{2}}_{n}A-A({{\theta}_{ns}}^{2}-{{\kappa}_{0}}^{2})]
\label{27}
\end{equation}
\begin{equation}
-{\kappa}_{0}{\tau}_{0}A=-uA[{\Omega}_{b}+{\tau}_{0}]+{\epsilon}[2{\kappa}_{0}+({\Omega}_{s}+{\tau}_{0}){\theta}_{ns}]A
\label{28}
\end{equation}
Where we have used the hypothesis of helical or screwed dynamos
where torsion and curvature coincides and are constants and equal to
${\tau}_{0}$ and ${\kappa}_{0}$. Simple algebraic manipulation of
these equations, in the limit of zero resistivity ,or ${\epsilon}=0$
reduce them to
\begin{equation}
{{\kappa}_{0}}^{2}+u{\kappa}_{0}+{\Omega}_{b}u=0 \label{29}
\end{equation}
which is a second order algebraic equation, and
\begin{equation}
{\partial}_{t}A=[2{{\kappa}_{0}}^{2}-{{\theta}_{ns}}^{2}]A
\label{30}
\end{equation}
\begin{equation}
{\partial}_{n}A={{\kappa}_{0}}A \label{31}
\end{equation}
The discriminant of the algebraic equation is assumed to vanish or
${\Delta}={u}^{2}-4{\Omega}_{b}u=0$ which yields the solution
$u=-4{\Omega}_{b}$. A simple solution of the system can now be
obtained by separation of variables $A(t,n)=H(n)T(t)$, substitution
of this product in the equations yields
\begin{equation}
A=A_{0}e^{[{\lambda}t-{{\kappa}_{0}}n]} \label{32}
\end{equation}
which by the relation between the fields B and A yields
\begin{equation}
B=A_{0}({\Omega}_{b}+{\tau}_{0})e^{[{\lambda}t-{{\kappa}_{0}}n]}
\label{33}
\end{equation}
where ${\lambda}:=[2{{\kappa}_{0}}^{2}-{{\theta}_{ns}}^{2}]$. Both
results shows that the filament magnetic field can only be
maintained in time when ${\lambda}>0$ or
$2{{\kappa}_{0}}^{2}>{{\theta}_{ns}}^{2}$ otherwise the magnetic
field decay very fast in time and the filament is actually an
antidynamo. Note that when the filament is isotropic in the sense
the frame only depends upon the arc length s then the filaments are
always dynamos since then ${\theta}_{ns}=0$.
\section{Topology bounds the  energy from magnetic helicity}
The magnetic energy of a divergence vector field on a compact planar
domain M reads
\begin{equation}
E_{B}=\frac{1}{8{\pi}}\int{B^{2}dV} \label{34}
\end{equation}
substitution of the formulas obtained in the last section yields
\begin{equation}
E_{B}=\frac{a^{2}}{8}\int{{A_{0}}^{2}[{{\Omega}_{b}+{\tau}_{0}}^{2}]ds}
\label{35}
\end{equation}
where we consider that the magnetic filaments leads constant
cross-section area as $S={\pi}a^{2}$. In terms of the vector
potential component $A_{n}$ expression (\ref{35}) reduces to
\begin{equation}
E_{B}=\frac{La^{2}}{8}[B^{2}]\label{36}
\end{equation}
where $L=\int{ds}$ is the length of the filament. Now the Arnold
theorem states that a divergence-free vector field ${\alpha}$
defined a compact manifold M , the helicity have an upper limit
exactly given by the modulus of the magnetic helity
\begin{equation}
H=\int{\vec{\alpha}.{\nabla}{\times}\vec{\alpha}dV} \label{37}
\end{equation}
where the topological bound is given by \cite{38}
\begin{equation}
E_{B}>C|H| \label{38}
\end{equation}
where C is a positive constant which depends upon the size and form
of the compact manifold. Let us now apply the Arnold's theorem to
our solution and analyze the implications for the anti-dynamo
problem. First of all we must compute the magnetic helicity in the
example given here. This yields
\begin{equation}
|H|={\pi}a^{2}L[{\Omega}_{b}+{\tau}_{0}]{A}^{2}e^{[{\lambda}t-{\kappa}_{0}n]}
\label{39}
\end{equation}
Thus to examine the comparison with the Arnold's theorem we have
\begin{equation}
C=\frac{E_{B}}{|H|}=\frac{\pi}{8}[{{\Omega}_{b}}+{\tau}_{0}]
\label{40}
\end{equation}
which is the Arnold's constant C which clearly fulfill the
geometrical requirements of Arnold's theorem. One must notice that
when the dynamo conditions discussed in the previous section is
fulfilled the Arnolds theorem is also obeyed, however when dynamos
existence is not possible Arnold's theorem seems to be violated and
an lower bound for the energy is not obtained anymore.
\section{Conclusions}
 While the existence of planar MHD solutions does not warrant the existence of dynamos (with the possible exception of the
 presence of fold of filaments) it is enough to
 warrant the existence of the amplification of magnetic fields. Unfortunately a simple way, though not the only, of getting fold
 of the filaments is by introducing Frenet torsion in the filaments but this would violate the condition of planarity.
 In conclusion, the investigation of kinematical anti-dynamos and dynamos filamentary MHD in generalised Frenet
 frame shows that is possible to test the Arnold's theorem against the dynamo conditions on Frenet curvature
 and topology comparing it with the magnetic helicity of streamlines. We pretend in near future to investigate this theorem
 for a more general class of fluids where the topological numbers of
 twist and writhe can be computed for other more complex filamentary geometries.

 \section*{Acknowledgements}
 Thanks are due to CNPq and UERJ for financial supports.


\newpage
\begin{thebibliography}{8}
\bibitem{1} V. Arnold and B. Khesin, Topological Methods in
Hydrodynamics, Applied Mathematics Sciences 125 (1991).
\bibitem{2} S. Childress and A. Gilbert, Stretch, Twist and Fold: The Fast Dynamo (1996)(Springer).
\bibitem{3} K. H. Moffatt, Nature 341 (1989).
\bibitem{4} J. Thiffault and A.H.Boozer, The Onset of Dissipation in the Kinematic Dynamo,Los Alamos arXiv:nlin.CD/0209042v1.L.C. Garcia de Andrade, Physics of Plasmas 13, 022309 (2006).
\bibitem{5} L. C. Garcia de Andrade, Curvature and Torsion effects
on carrying currents twisted solar loops, (2006) Phys of Plasmas nov
issue. L.C.Garcia de Andrade, Phys Scripta 73 (2006).
\bibitem{6} A. Schekochihin, S. Cowley,J. Maron and L. Malyshkin, Phys Rev E (2001) .
\bibitem{7} L.C.Garcia de Andrade, Phys Scripta 73 (2006).
\bibitem{8} B. Khesin, Topology Bounds Energy, in reference $1$.
\end{thebibliography}
\end{document}